\documentclass[preprint,showpacs,preprintnumbers,aps]{revtex4}%
\usepackage{amsmath}
\usepackage{graphicx}
\usepackage{dcolumn}
\usepackage{bm}
\usepackage{amsfonts}
\usepackage{amssymb}%
\begin{document}
\title{Monopole and quadrupole polarization effects on the $\alpha$-particle
description of $^{8}$Be}
\author{A. Sytcheva}
\author{F. Arickx}
\author{J. Broeckhove}
\affiliation{University of Antwerp, Antwerp, B2020, Belgium}
\author{V. S. Vasilevsky}
\affiliation{Bogolyubov Institute for Theoretical Physics, Kiev, 03143, Ukraine}

\begin{abstract}
We investigate the effect of monopole and quadrupole modes on the elastic
$\alpha-\alpha$ resonance structure of $^{8}$Be. To this end we make a fully
microscopic coupled channels calculation with three coupled channels, using
the Algebraic Model. The continuum spectrum and wave functions are analyzed in
terms of the individual channels to understand the nature of the resonances.
It is shown that both monopole and quadrupole modes have a non-negligible
effect on the resonances in the $\alpha-\alpha$\ continuum.

\end{abstract}
\date{{\today}}
\pacs{21.60.Gx, 21.60.Ev, 25.55.ci, 24.30.Gd.}
\pacs{21.60.Gx, 21.60.Ev, 25.55.ci, 24.30.Gd.}
\pacs{21.60.Gx, 21.60.Ev, 25.55.ci, 24.30.Gd.}
\pacs{21.60.Gx, 21.60.Ev, 25.55.ci, 24.30.Gd.}
\pacs{21.60.Gx, 21.60.Ev, 25.55.ci, 24.30.Gd.}
\pacs{21.60.Gx, 21.60.Ev, 25.55.ci, 24.30.Gd.}
\pacs{21.60.Gx, 21.60.Ev, 25.55.ci, 24.30.Gd.}
\pacs{21.60.Gx, 21.60.Ev, 25.55.ci, 24.30.Gd.}
\pacs{21.60.Gx, 21.60.Ev, 25.55.ci, 24.30.Gd.}
\pacs{21.60.Gx, 21.60.Ev, 25.55.ci, 24.30.Gd.}
\pacs{21.60.Gx, 21.60.Ev, 25.55.ci, 24.30.Gd.}
\pacs{21.60.Gx, 21.60.Ev, 25.55.ci, 24.30.Gd.}
\pacs{21.60.Gx, 21.60.Ev, 25.55.ci, 24.30.Gd.}
\pacs{21.60.Gx, 21.60.Ev, 25.55.ci, 24.30.Gd.}
\pacs{21.60.Gx, 21.60.Ev, 25.55.ci, 24.30.Gd.}
\pacs{21.60.Gx, 21.60.Ev, 25.55.ci, 24.30.Gd.}
\pacs{21.60.Gx, 21.60.Ev, 25.55.ci, 24.30.Gd.}
\pacs{21.60.Gx, 21.60.Ev, 25.55.ci, 24.30.Gd.}
\pacs{21.60.Gx, 21.60.Ev, 25.55.ci, 24.30.Gd.}
\pacs{21.60.Gx, 21.60.Ev, 25.55.ci, 24.30.Gd.}
\pacs{21.60.Gx, 21.60.Ev, 25.55.ci, 24.30.Gd.}
\pacs{21.60.Gx, 21.60.Ev, 25.55.ci, 24.30.Gd.}
\pacs{21.60.Gx, 21.60.Ev, 25.55.ci, 24.30.Gd.}
\pacs{21.60.Gx, 21.60.Ev, 25.55.ci, 24.30.Gd.}
\pacs{21.60.Gx, 21.60.Ev, 25.55.ci, 24.30.Gd.}
\pacs{21.60.Gx, 21.60.Ev, 25.55.ci, 24.30.Gd.}
\pacs{21.60.Gx, 21.60.Ev, 25.55.ci, 24.30.Gd.}
\pacs{21.60.Gx, 21.60.Ev, 25.55.ci, 24.30.Gd.}
\maketitle

\section{Introduction}

$^{8}$Be is known to be a strongly clustered nucleus that appears through
relatively short-lived resonances just above the $\alpha-\alpha$ scattering
threshold. A low-lying rotational band is experimentally apparent and suggests
a strong deformation of the 8-particle resonance system. A systematic survey
of the full spectrum of $^{8}$Be, including a review of available theoretical
and experimental results, has been made in the work of Ajzenberg-Selove
\cite{ar:ajzsel-NPA490-1} and is now available\ in a revised version
\cite{ar:tilley2004}. Bacher et al. \cite{ar:Bacher-PRL29-1331} have reported
partial phase shifts in the $\alpha-\alpha$ collisions for even states up to
$L=6$ in the range up to 35 MeV excitation energy. Together with
\cite{AR:darriulat-PR-137-315}, these experiments reveal no resonance states
with measurable widths above 25 up to 50 MeV in the elastic $\alpha-\alpha$
channel. Arena et al. \cite{AR:ostashko-JPG-20-1973} point out the need of
including inelastic channels such as $^{4}$He($^{4}$He,d)$^{6}$Li, $^{4}%
$He($^{4}$He,n)$^{7}$Be, and $^{4} $He($^{4}$He,p)$^{7}$Li if one wants to
find states of high excitation energy. These authors report the possible
existence of highly excited\ $^{8}$Be levels for $L=6$ \ and $L=12$ at about
41 MeV and $E_{x}=43$, and around 50 MeV for $L=2$ up to $L=10$.

Because of the strong experimental evidence for predominance of cluster
structure in $^{8}$Be, many theoretical approaches based on cluster structures
have been considered. Microscopic cluster models are known to provide valuable
information about the structure of light nuclei \cite{kn:wildtang77},
\cite{kn:tang81} and in particular of $^{8}$Be. The Resonating Group Method
(RGM) has often been used \cite{ar:Tang78}, \cite{ar:wheeler-PR52-1083},
\cite{ar:HillWheeler53}, \cite{ar:saito77}. The low-lying rotational structure
of $^{8}$Be, both in position and width, is reproduced by elastic
$\alpha-\alpha$ scattering calculations with effective interactions. The
Coulomb interaction plays an important role in the correct position and width
of these states, in particular for $L=0$ groundstate of $^{8}$Be
\cite{ar:arai-PRC60-064315}, \cite{AR:humblet-NPA-638-714}.

Collective $A$-particle deformation models for light nuclei have been under
discussion for several decades. A meaningful classification scheme has been
derived for such models through the irreducible representations of the
non-compact Sp(2, R) group. A good description of the low-energy spectra of
light nuclei \cite{ar:caurierarickx-NPA398-467}, and more specifically of the
rotational structure of $^{8}$Be \cite{AR:deum-NPA-252-416} has been obtained
within these models. It has also been demonstrated that the quadrupole Sp(2,R)
model of $^{8}$Be and the $\alpha$-particle description have an important
overlap, and thus are complementary in the description of this nucleus
\cite{AR:arickx-NPA-284-264}, a conclusion confirmed in
\cite{ar:Rowe-RPP48-1419}, \cite{ar:suzuki-PTP75-1377},
\cite{ar:HechtBraunsch-NPA295-34}. It seems therefore appropriate to study the
coexistence and competition between collectivity and clustering in light
nuclei through a combined approach. Filippov et al. \cite{ar:FilVasNes},
\cite{ar:fil-NuovCim89}, \cite{kn:VV86collresE} have already investigated
cluster-monopole and cluster-quadrupole descriptions of $^{8}$Be.

The deformation aspect in $^{8}$Be has been studied within a cluster approach
by introducing a quadrupolar polarization of the $\alpha$-particles
\cite{ar:deumens:NPA423}, or a monopolar distortion of the $\alpha$-particles
\cite{AR:kruglanski-PRC-45-1321}.

In this work we propose a model in which the $\alpha$-particle description and
the collective (8-particle) quadrupole and monopole modes are coupled. We
consider a scattering approach within an energy range in which the $\alpha$
cluster channel is open, and both collective channels are closed, thus
limiting ourselves to the elastic $\alpha-\alpha$ cluster decay. The effects
of the collective channels will then only be apparent in the compound system
during resonance lifetimes.

We implement our coupled channels approach within the Modified J-Matrix Method
(MJM) \cite{ar:PRL88-10404}, \cite{ar:MJM4shortrange-jphysA-7769} also known
as the Algebraic Model \cite{ar:AM-AJP},\cite{kn:VA-PhysRev}. It determines an
approximate solution of the Schr\"{o}dinger equation in terms of
square-integrable bases, and maps both scattering or bound-state boundary
conditions from configuration to the space of basis expansion coefficients. As
such it allows for a simultaneous treatment of open and closed channels. The
MJM is an extension of the J-Matrix Method (JM) \cite{ar:PRA-9-1974HY(JMM)}
using an oscillator basis. The MJM allows one to treat long-range
interactions, including the Coulomb potential, in that basis. It provides
convergence in terms of number of basis functions with reasonable basis sizes,
which is important because calculation of Hamiltonian matrix elements is the
bulk of the computational load of the method.

The multi-channel approach of this work allows for a clear-cut analysis of the
elastic $\alpha-\alpha$ phase shifts and corresponding wave functions in terms
of the contribution from the individual channels, leading to a physical
interpretation of the resonances. It indicates the importance of collective
degrees of freedom in the compound system.

Our approach is most suited to handle two-body Gaussian interactions, mainly
because of the oscillator expansion in the MJM. We consider two-body
potentials such as the one proposed in \cite{ar:volkov-NP74-33} and determined
within a Hartree-Fock approximation, and the ones from
\cite{ar:tang-NPA286-53}, \cite{ar:tonabe-PTP53-677} determined within an RGM
approach, to calculate the $^{8}$Be spectrum, and check the validity of our conclusions.

The paper is organized as follows. In section II we elaborate on the combined
cluster-collective model description for $^{8}$Be and formulate the
multi-channel MJM scattering approach. In Section III we discuss the numerical
application of the MJM and present the results of the three-channel
calculation. Section IV is devoted to the analysis of the results of the
previous section in terms of contributions from the individual channels.
Concluding remarks are presented in Section V.

\section{A Coupled-Channels Cluster-Collective Modified J-matrix Approach}

The Modified J-Matrix Method \cite{ar:PRL88-10404},
\cite{ar:MJM4shortrange-jphysA-7769} also referred to as the Algebraic Version
of the RGM \cite{kn:VA-PhysRev}, has become a well-tested approach for nuclear
structure calculations involving multi-channel cluster and/or collective
descriptions for light nuclei. The application of the MJM is based on an
expansion in terms of oscillator basis states in the respective collective
coordinates (intercluster distance, monopole radius, quadrupole deformation,
...). We refer to the papers of Vasilevsky et al. \cite{ar:JPhysG18-1227},
\cite{ar:vasnesarbr-PAN1997}, \cite{ar:vas-PRC63-034606},
\cite{ar:vas-PRC63-034607} for detailed properties of the individual channel
wave functions, and the multi-channel formulation of the MJM
\cite{ar:threecluster-alh} with non-orthogonal bases.

The model considered here for $^{8}$Be consists of a wave function containing
three structure components distinguished by a specific collective coordinate.
These three components represent the $\alpha-\alpha$ cluster, the Sp(2,R)
quadrupole and the Sp(2,R) monopole modes%
\begin{equation}
\Psi=\Psi^{C}+\Psi^{Q}+\Psi^{M}. \label{eq:wf}%
\end{equation}

The structure of a single cluster is described by a wave function $\Psi
_{i}\left(  \alpha_{i}\right)  $%
\begin{equation}
\Psi_{i}\left(  \alpha_{i}\right)  =\Psi_{i}\left(  \mathbf{q}_{1}^{\left(
i\right)  },\mathbf{q}_{2}^{\left(  i\right)  },\mathbf{q}_{3}^{\left(
i\right)  }\right)  ,\quad(i=1,2)
\end{equation}
centered around its centre of mass $\mathbf{R}_{i}$.

A two-cluster wave function can then be written as%
\begin{equation}
\Psi^{C}\left(  \mathbf{q}_{1},..,\mathbf{q}_{7}\right)  =\mathcal{A}\left[
\Psi_{1}\left(  \alpha_{1}\right)  \ \Psi_{2}\left(  \alpha_{2}\right)
\ \Psi_{R}\left(  \mathbf{r}\right)  \right]  ,
\end{equation}
where $\mathcal{A}$ stands for the antisymmetrization operator over all $8$
particles, and $\Psi_{R}\left(  \mathbf{r}\right)  $ represents the relative
motion of both clusters, $\mathbf{r}$\ being the corresponding Jacobi coordinate.

To limit the computational complexity of the problem, the cluster wave
functions are frozen, and constructed as Slater determinants of harmonic
oscillator ($0s$)-states, corresponding to the groundstate shell-model
configuration of the cluster. The $\Psi_{R}\left(  \mathbf{r}\right)  $ wave
function for the relative motion will be represented by an expansion in terms
of an oscillator basis in $\mathbf{r}$. As the cluster states are frozen and
built of ($0s$)-orbitals, the quantum numbers reduce to those of the
inter-cluster wave function only. The set of quantum numbers is unambiguously
defined and is obtained from the reduction of the symmetry group $U(3)\supset
O(3)$ of the one-dimensional oscillator. This reduction provides the quantum
numbers $n$ for the radial excitation, and $L,M$ for the angular momentum of
the two-cluster system. The two-cluster wave function can be decomposed as%
\begin{align}
\Psi_{LM}^{C}  &  =\mathcal{A}\left[  \Psi_{1}\left(  \alpha_{1}\right)
\ \Psi_{2}\left(  \alpha_{2}\right)  \ \Psi_{LM}\left(  \mathbf{r}\right)
\right] \label{eq:ClExpansion}\\
&  =\sum_{n}c_{nL}^{C}\mathcal{A}\left[  \Psi_{1}\left(  \alpha_{1}\right)
\ \Psi_{2}\left(  \alpha_{2}\right)  \ \phi_{nLM}\left(  \mathbf{r}\right)
\right]  =\sum_{n}c_{nL}^{C}\psi_{nLM}^{C},\nonumber
\end{align}
where the $\phi_{nLM}\left(  \mathbf{r}\right)  $ are the three-dimensional
harmonic oscillator states. The oscillator parameter $b$ is the same for both
the individual particle ($0s$) states, and the expansion for the relative motion.

The oscillator decomposition of collective Sp(2,R) quadrupole and monopole
components are most easily introduced through the standard step operators%
\begin{align}
\Psi_{LM}^{Q}  &  =\sum_{n}c_{nLM}^{Q}N_{LM}^{(Q)}P_{LM}\left[  A_{Q}%
^{+}\right]  ^{n}\Phi_{0}=\sum_{n}c_{nLM}^{Q}\psi_{nLM}^{Q}%
;\label{eq:QExpansion}\\
\Psi_{LM}^{M}  &  =\sum_{n}c_{nL0}^{M}N_{LM}^{(M)}\left[  A_{M}^{+}\right]
^{n}P_{LM}\Phi_{0}=\sum_{n}c_{nLM}^{M}\psi_{nLM}^{M}, \label{eq:MExpansion}%
\end{align}
where $\Phi_{0}$ is a $0\hbar\omega$ shell-model vacuum state. For $^{8}$Be
this is a Slater determinant with a $(0s)^{4}(0p)^{4}=(000)^{4}(001)^{4}$
configuration (in Cartesian $(n_{x},n_{y},n_{z})$ oscillator notation) and it
has SU(3) $(\lambda,\mu)=(40)$ classification. The $P_{LM}$ stands for the
traditional angular momentum projection operator and $N_{LM}^{(M)}$,
$N_{LM}^{(Q)}$ are norm factors. The (translationally invariant) step
operators $A_{\mu\nu}^{+}$ in a Cartesian notation $(\mu,\nu=x,y,z)$ are
written, in terms of the standard harmonic oscillator creation operators
$a_{\mu}^{+}(i)$ for particle $i$:%
\begin{equation}
A_{\mu\nu}^{+}=\sum_{i=1}^{A}a_{\mu}^{+}(i)a_{\nu}^{+}(i)-\frac{1}{A}%
\sum_{i,j=1}^{A}a_{\mu}^{+}(i)a_{\nu}^{+}(j)
\end{equation}
so that%
\begin{align}
A_{Q}^{+}  &  =A_{zz}^{+};\nonumber\\
A_{M}^{+}  &  =A_{xx}^{+}+A_{yy}^{+}+A_{zz}^{+}.
\end{align}

The projection after excitation for the quadrupole mode in
(\ref{eq:QExpansion}) is necessary because the step operator $A_{Q}^{+}$
contains both $L=0$ and $L=2$ components. In the monopole case, the projection
operator commutes with the step operators. A single projection of the
$0\hbar\omega$ state, which contains $L=0,2$ and $4$ components suffices. Thus
the monopole mode contributes only to those $L$-subspaces. Contrary to the
cluster and quadrupole modes, this limits the monopole Hilbert space to the
latter angular momenta. We will therefore limit ourselves to $L=0,2$ and $4$
for all modes throughout this work.

We use the same oscillator parameter $b$ for $\Psi_{LM}^{Q}$ and $\Psi
_{LM}^{M}$ functions as we do for the cluster wave function $\Psi_{LM}^{C}$.
This makes the calculation of overlap and \ Hamiltonian matrix elements
significantly easier.

One has of course orthogonality with respect to the quantum numbers $(n,L,M)$,
but not with respect to the channel label:%
\begin{equation}
\left\langle \psi_{nLM}^{\tau}|\psi_{n^{\prime}L^{\prime}M^{\prime}}%
^{\upsilon}\right\rangle \sim\ \delta_{n,n^{\prime}}\delta_{L,L^{\prime}%
}\delta_{M,M^{\prime}},
\end{equation}
where $\tau$ and $\upsilon$ stand for the cluster ($C$), monopole ($M$) and
quadrupole ($Q$)channels. In particular it is well known
\cite{AR:arickx-NPA-284-264} that with a common choice of overall $b$ the
$n=0$ basis states of all modes are identical:
\begin{equation}
\psi_{0LM}^{C}=\psi_{0LM}^{Q}=\psi_{0LM}^{M}. \label{eq:SameFirstBasisState}%
\end{equation}

This means that care should be taken in the interpretation of the results when
attributing an effect to some channel or other.

The calculation of overlap and Hamiltonian matrix elements is most easily
performed by considering Gaussian-type generating functions for the three
oscillator expansions, which are (up to $L$-projection)%
\begin{align}
\psi^{C}(\mathbf{R})  &  =\mathcal{A}\left[  \Psi_{1}\left(  A_{1}\right)
\ \Psi_{2}\left(  A_{2}\right)  \ \phi\left(  \mathbf{r|R}\right)  \right]
\nonumber\\
&  =\mathcal{A}\left[  \Psi_{1}\left(  A_{1}\right)  \ \Psi_{2}\left(
A_{2}\right)  \ \exp\left\{  -\frac{1}{2b^{2}}\mathbf{r}^{2}+\frac{\sqrt{2}%
}{b^{2}}\left(  \mathbf{Rr}\right)  -\frac{1}{2b^{2}}\mathbf{R}^{2}\right\}
\right] \nonumber\\
&  =\mathcal{A}\left[  \Psi_{1}\left(  A_{1}\right)  \ \Psi_{2}\left(
A_{2}\right)  \ \exp\left\{  \left(  \mathbf{R\bullet A}_{C}^{+}\right)
\right\}  \exp\left\{  -\frac{1}{2b^{2}}\mathbf{r}^{2}\right\}  \right]
;\nonumber\\
\psi^{Q}(\epsilon)  &  =\exp\{\epsilon A_{Q}^{+}\}\Phi_{0};\nonumber\\
\psi^{M}(\nu)  &  =\exp\{\nu A_{M}^{+}\}\Phi_{0},
\end{align}
where we introduced the step operator $\mathbf{A}_{C}^{+}=\mathbf{a}%
^{+}(\mathbf{r})$ for the oscillator decomposition of the cluster wave function.

Because of the Gaussian nature of the generating functions, matrix elements
for the overlaps and Gaussian two-body operators can be calculated in a
straightforward way. From these the matrix elements in the discrete $(n,L,M)$
basis can then be obtained by e.g. recurrence techniques \cite{kn:VA-PhysRev},
\cite{ar:threecluster-alh}.

By substituting the expansions (\ref{eq:ClExpansion}, \ref{eq:QExpansion},
\ref{eq:MExpansion}) as an ansatz for the solution (\ref{eq:wf}) in the
Schr\"{o}dinger equation, the latter reduces to an infinite matrix equation to
be solved for the coefficients $c_{nL}^{C}$, $c_{nL}^{Q}$ and $c_{nL}^{M}$:%
\begin{equation}
\sum_{\tau^{\prime}}\sum_{m}\left\langle nL,\tau\left\vert \hat{H}%
-E\right\vert mL,\tau^{\prime}\right\rangle c_{mL}^{\tau^{\prime}}=0.
\end{equation}

We solve this equation by considering the Modified J-Matrix approach, which
was formulated in terms of an oscillator decomposition of the trial solution,
and provides fast convergence in a finite subset of the model subspace. In the
J-Matrix approach the boundary conditions (for scattering as well as
bound-state solutions) are translated from coordinate into the space of basis
expansion coefficients, and asymptotic solutions can be obtained from a
three-term recurrence relation for the expansion coefficients for high $n$.
Considering the coefficients to equal the asymptotic values from a given $n=N$
on, and imposing a matching condition between the interaction region and
asymptotic regions, only a reasonably sized matrix equation remains to be
solved. In the Modified J-Matrix approach asymptotic contributions for the
potential behavior, in particular for the Coulomb term, are taken into account
in a semi-classical way through a modified recurrence equation. This was shown
to reduce the size $N$ of the\ remaining matrix equation drastically.

For the scattering boundary condition the asymptotic regular $(c_{nL}^{(+)})$
and irregular $(c_{nL}^{(-)})$ solutions are obtained so that the asymptotic
expansion coefficients of the solution can be written as%
\begin{equation}
c_{nL}^{as}\rightarrow c_{nL}^{(-)}(kR_{n,l})-Sc_{nL}^{(+)}(kR_{n,l}),\qquad
n\rightarrow\infty, \label{eq:openAsymp}%
\end{equation}
where $S$ stands for the $S$-matrix and reflects the matching condition. It is
to be determined by solving the remaining matrix equation. $k=\sqrt
{2mE/\hbar^{2}}$ is the momentum corresponding to energy $E$, and $R_{n,l}$
are the oscillator turning points for the channel under consideration. In a
single-channel approach $S$ is a scalar quantity related to the phase-shift only.

For the bound-state boundary condition only the exponentially decaying
solution can be retained, and its expansion coefficients are%
\begin{equation}
c_{nL}^{as}\rightarrow\exp(-\varkappa R_{n,l})/\sqrt{R_{n,l}},\qquad
n\rightarrow\infty\label{eq:boundAsymp}%
\end{equation}
with $\varkappa=\sqrt{2m\left\vert E\right\vert /\hbar^{2}}$. In the equations
(\ref{eq:openAsymp}) and (\ref{eq:boundAsymp}) the energy $E$ is determined
with respect to threshold of the corresponding channels.

In our approach we have three different channels, so a multi-channel MJM
formulation is necessary. We take the same form as in
\cite{ar:vas-PRC63-034606} and write
\begin{equation}
c_{nL}^{\tau}=c_{nL}^{(0)\tau}+\delta_{\upsilon\tau}c_{nL}^{(-)\tau
}-S_{\upsilon\tau}c_{nL}^{(+)\tau} \label{eq:ChannelCoeffs}%
\end{equation}
with $\upsilon$ the entrance channel, and $\tau$ any other coupled channel.
Substitution of (\ref{eq:ChannelCoeffs}) in the matrix form of the
Schr\"{o}dinger equation leads to%

\begin{align}
\sum_{\tau^{\prime}}\sum_{m<N}\left\langle nL,\tau\left\vert \hat
{H}-E\right\vert mL,\tau^{\prime}\right\rangle c_{mL}^{(0)\tau^{\prime}}-  &
\sum_{\tau^{\prime}}S_{\upsilon\tau^{\prime}}\left[  \beta_{0}^{(+)\tau
^{\prime}}\ \delta_{n,0}\ \delta_{\tau^{\prime}\tau}+V_{n}^{\left(  +\right)
\tau\tau^{\prime}}\right] \nonumber\\
&  =-\beta_{0}^{(-)\tau}\ \delta_{n,0}\ \delta_{\upsilon\tau}-V_{n}^{\left(
-\right)  \tau\upsilon} \label{eq:FinalMJMeqs}%
\end{align}
and
\begin{equation}
V_{n}^{(\pm)\tau\tau^{\prime}}=\sum_{m=0}^{\infty}\left\langle nL,\tau
\left\vert \hat{V}\right\vert mL,\tau^{\prime}\right\rangle \ c_{m}^{(\pm
)\tau^{\prime}}, \label{eq:DynamicCoeff}%
\end{equation}
where $\hat{V}$\ stands for the two-body interaction and $\beta_{0}$ accounts
for the traditional regularization of the irregular asymptotic solution (see
for instance \cite{kn:VA-PhysRev} or \cite{ar:vas-PRC63-034606}).

This system of equations should then be solved for the residual coefficients
$c_{nL}^{\left(  0\right)  \tau}$ and the $S$-matrix elements $S_{\tau
\tau\prime}$. We consider in equations (\ref{eq:FinalMJMeqs}%
-\ref{eq:DynamicCoeff}) a near-interaction region with $n<N$ and a
far-interaction region with $n\geq N$. The choice of $N$ is such that one can
expect the residual expansion coefficients $\left\{  c_{nL}^{(0)\tau}\right\}
$ to be negligibly small in the far-interaction region. The total number of
equations for a given entrance channel $\upsilon$ then equals to
$N_{ch}\left(  N+1\right)  $, and solving the set of equations by traditional
numerical linear algebra leads to the $N_{ch}N$ residual coefficients
$\left\{  c_{n}^{\left(  0\right)  \tau}\text{; }\tau=C,Q,M\text{;
}n=0..N-1\right\}  $ and $N_{ch}$ $S$-matrix elements $\left\{  S_{\upsilon
\tau}\text{; }\tau=C,Q,M\right\}  $. The set of equations has to be solved for
all $N_{ch}$ entrance channels. A final parameter of the calculation concerns
the summation in (\ref{eq:DynamicCoeff}). Because the potential matrix
elements decrease rapidly when $\left\vert n-m\right\vert $ gets large, we can
truncate this sum at some $M>N$.

In this paper we limit ourselves to the situation in which only the cluster
channel is open, so that $\upsilon=C$, because this is the dominant channel
for $^{8}$Be in our model description. The boundary conditions
(\ref{eq:openAsymp}) are appropriate for an open channel, and are therefore
used for the cluster channel; conditions (\ref{eq:boundAsymp}) are appropriate
for a closed channel and are thus applied to the monopole and quadrupole
channels. This choice limits the energy range between the monopole threshold
at 0 MeV (all 8 particles infinitely apart) and the cluster threshold. The
quadrupole threshold is even higher, because of the forced polarization condition.

\section{Numerical Application and Results}

In a microscopic calculation the choice of the effective nucleon-nucleon (NN)
interaction remains a crucial point. We limit ourselves to effective NN
interactions of Gaussian form, which lead to a straightforward evaluation of
matrix elements in the cluster-collective model space. One well-known example
is Volkov \cite{ar:volkov-NP74-33} force, which was essentially determined and
used within a Hartree-Fock context. This force binds both the deuteron triplet
as well as the dinucleon singlet. Gaussian forces that discriminate between
the deuteron triplet and dinucleon triplet are the Minnesota
\cite{ar:tang-NPA286-53} and the Hasegawa-Nagata \cite{ar:hasegawa-PTP38-118},
\cite{ar:hasegawa-PTP45-1786} potentials. A modified version of the latter was
proposed in \cite{ar:tonabe-PTP53-677}. These interactions were considered and
tested in $\alpha-N$ and $\alpha-\alpha$ RGM scattering calculations.

In the current calculations we consider the Volkov (V1), Modified
Hasegawa-Nagata (MHN) and Minnesota (Mi) forces. We include the Coulomb
interaction which is necessary to produce the $L=0$ ground-state as a narrow
resonance just above the $\alpha-\alpha$ threshold. The parameters are chosen
to reproduce both the ground state energy and size of $^{4}$He. The Majorana
exchange part accounts for nuclear matter properties. It does not influence
the ground state energy and size of $^{4}$He, but affects the deformation in
$p$-shell nuclei significantly \cite{ar:tang-NPA286-53}.

To obtain the phase-shifts for elastic $\alpha-\alpha$ scattering we solve
(\ref{eq:FinalMJMeqs}), and determine resonance positions and widths in the
usual numerical way through
\begin{equation}
\frac{d^{2}\delta_{l}}{dE^{2}}=0\Longrightarrow E_{r},\quad\Gamma=2\left(
\left.  \frac{d\delta_{l}}{dE}\right\vert _{E_{r}}\right)  ^{-1}.
\label{eq:phshanalysis}%
\end{equation}

We fix the common oscillator parameter $b$ for the C, Q and M expansion bases
for each of the potentials so as to optimize an acceptable $\alpha-\alpha$
threshold. These values can be found in Table \ref{tab:Forces}. We have also
slightly modified the Majorana parameters of the Minnesota and modified
Hasegawa-Nagata potentials to reproduce comparable values for the lowest
$L=0$, $2$ and $4$ resonances forming the ground state rotational band in
$^{8}$Be; the values for the modified Majorana parameters are also shown in
Table \ref{tab:Forces}. Also shown in the table are the respective threshold
energies for the three channels, which amounts to twice the binding energy of
the $\alpha$-particle for the cluster channel, zero energy (all 8 particles
apart) for the monopole channel, and a positive value for the quadrupole
channel (all 8 particles apart under the quadrupole deformation restriction).

\begin{table}[ptb]
\centering
\begin{tabular}
[c]{|l|c|c|c|}\hline
& Mi & V1 & MHN\\\hline
Original Majorana parameter & {0.52} & {0.6} & {0.39}\\\hline
Modified Majorana parameter & {0.57} & {0.6} & {0.43}\\\hline
Oscillator parameter $b$ (fm) & {1.28} & {1.37} & {1.32}\\\hline
$E_{\text{g.s.}}$of $^{4}$He, MeV & {-24.69} & {-27.09} & {-29.01}\\\hline
$E_{th}(C)$, MeV & {-49.37} & -54.17 & -58.02\\\hline
$E_{th}(M)$, MeV & {0.0} & {0.0} & {0.0}\\\hline
$E_{th}(Q)$, MeV & {39.22} & {23.16} & {25.28}\\\hline
\end{tabular}
\caption{Parameters used for Minnesota (Mi), Volkov (V1) and modified
Hasegawa-Nagata (MHN) potentials and corresponding cluster and quadrupole
threshold energies with respect to the monopole break-up energy.}%
\label{tab:Forces}%
\end{table}

\begin{figure}
[ptb]
\begin{center}
\includegraphics[
natheight=9.639200in,
natwidth=7.242800in,
height=20.1079cm,
width=15.1281cm
]%
{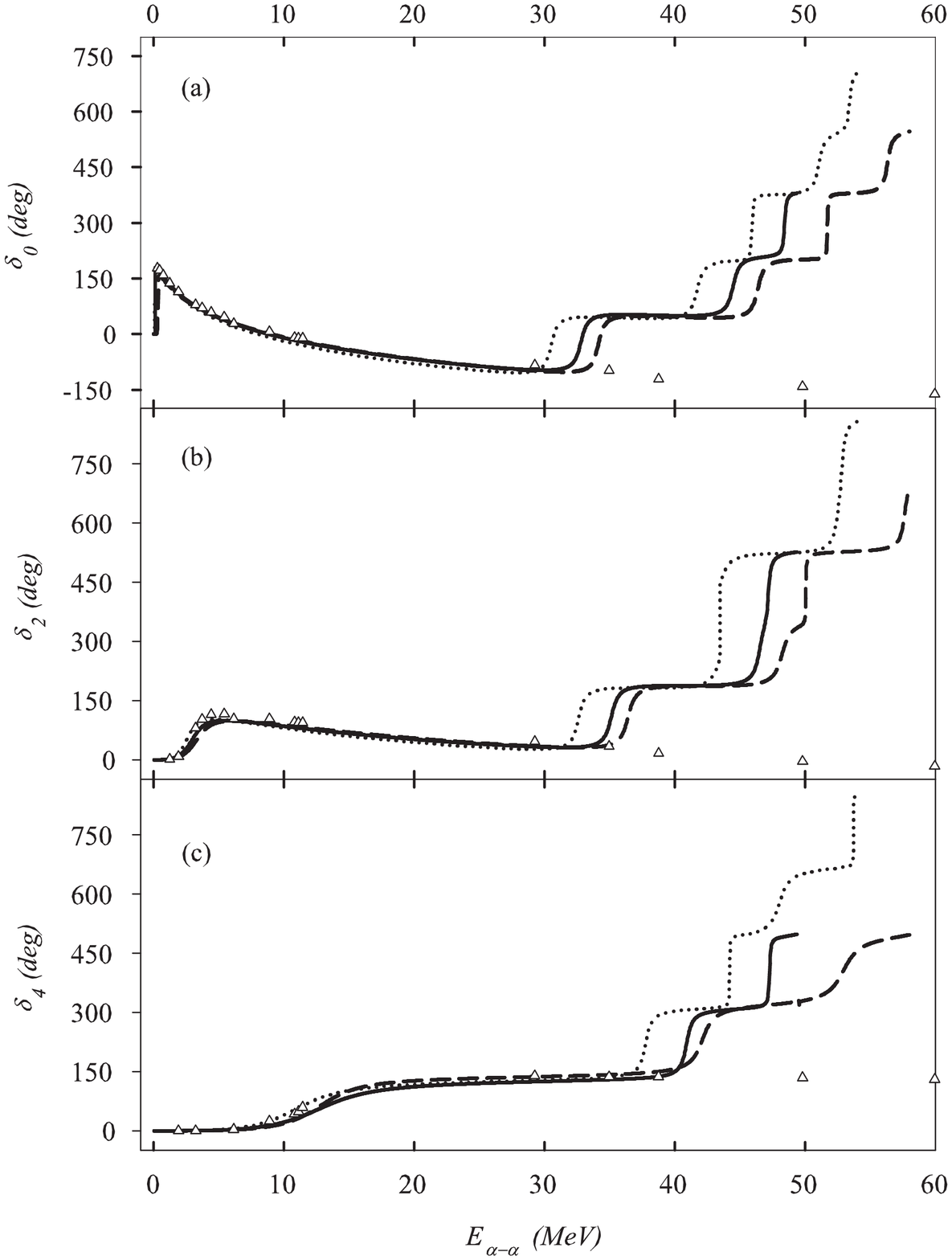}%
\caption{Phase shifts obtained in the multi-channel MJM approach with Minnesota (solid), Volkov (dotted) or MHN (dashed) forses for $	L=0$ (a), $L=2$ (b) and $L=4$ (c). Triangles indicate experimental data from \cite{ar:Bacher-PRL29-1331}, \cite{AR:darriulat-PR-137-315}. $E_{\alpha \alpha}$ is the c.m. energy with respect to the cluster threshold.}%
\label{f:all-shifts}%
\end{center}
\end{figure}

In Figure \ref{f:all-shifts} we show the elastic $\alpha$-$\alpha$ scattering
$L=0$, $2$ and $4$ phase shifts within the energy region between the $\alpha
$-$\alpha$ and monopole thresholds as obtained from the full MJM calculations,
including experimental data from \cite{ar:Bacher-PRL29-1331} and
\cite{AR:darriulat-PR-137-315}. The parameters of the calculation were chosen
for convergent phase shifts and, with $N=50$ in (\ref{eq:FinalMJMeqs}) for the
near interaction region, and $M=100$ for truncating the sum in
(\ref{eq:DynamicCoeff}) for all channels. The common states in the different
channels for $n=0$ have been taken into account properly, and forbidden Pauli
states are explicitly removed from the calculation.

\begin{figure}
[ptb]
\begin{center}
\includegraphics[
natheight=10.989200in,
natwidth=8.496800in,
height=11.1479cm,
width=15.6795cm
]%
{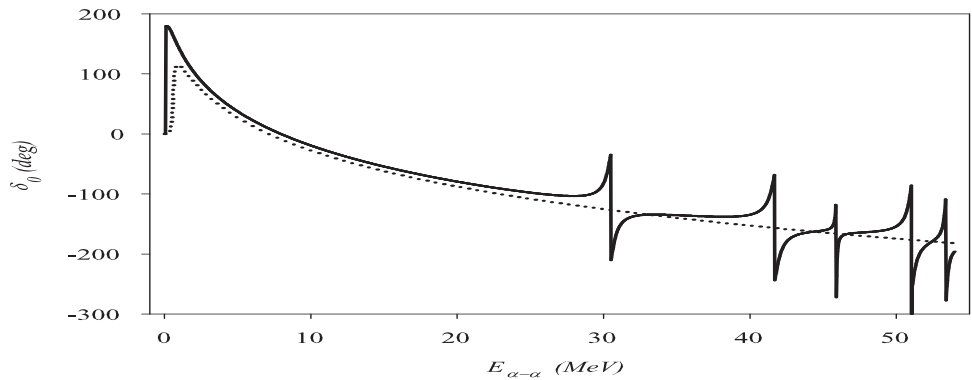}%
\caption{$L=0$ phase shifts for the Volkov potential in the one-channel cluster (dotted) and three-channel cluster-collective (solid line) approach.
$E_{\alpha-\alpha} $ is the c.m. energy with respect to the cluster threshold.}%
\label{f:elvsinel}%
\end{center}
\end{figure}

One immediately recognizes from Figure \ref{f:all-shifts} the low-lying
rotational structure of $^{8}$Be. A rich resonance structure beyond 30 MeV
above the cluster threshold appears through the coupling with the collective
channels. This is made clear in Figure \ref{f:elvsinel} where the pure cluster
phase shifts are compared to those of the fully coupled ones for V1 and $L=0$.
The effects are qualitatively well reproduced by all potentials considered,
implying that the effect is a genuine one, independent of specific choice of
NN-interaction. Table \ref{tab:Resonances} lists the resonance parameters
obtained from the phase shifts in Figure \ref{f:all-shifts} by applying
(\ref{eq:phshanalysis}).

\begin{table}[ptb]
\centering
\begin{tabular}
[c]{|l||c|c||c|c||c|r||}\hline
& \multicolumn{2}{c||}{Mi} & \multicolumn{2}{c||}{V1} &
\multicolumn{2}{c||}{MHN}\\\hline
& $E_{r}$ & $\Gamma$ & $E_{r}$ & $\Gamma$ & $E_{r}$ & $\Gamma$\\\hline
$0_{1}^{+}$ & {0.120} & {195} & {0.090} & {1910} & {0.085} & {0.2}\\\hline
$0_{2}^{+}$ & {32.78} & {823} & {30.51} & {761} & {33.78} & {827}\\\hline
$0_{3}^{+}$ & {44.45} & {762} & {41.72} & {843} & {46.04} & {876}\\\hline
$0_{4}^{+}$ & {48.41} & {208} & {45.92} & {95} & {51.42} & {76}\\\hline
$0_{5}^{+}$ & x & x & 51.06 & 616 & 56.08 & 618\\\hline
$0_{6}^{+}$ & x & x & 53.41 & 285 & x & x\\\hline
$2_{1}^{+}$ & {2.94} & {2126} & {2.56} & {1679} & {2.98} & {1712}\\\hline
$2_{2}^{+}$ & {35.13} & {8418} & {32.59} & {7977} & {36.08} & {1052}\\\hline
$2_{3}^{+}$ & {46.51} & {636} & {43.48} & {949} & {47.90} & {1128}\\\hline
$2_{4}^{+}$ & {47.17} & {235} & {44.48} & {0.2} & {49.69} & {80}\\\hline
$2_{5}^{+}$ & x & x & 52.84 & 250 & 57.40 & 554\\\hline
$4_{1}^{+}$ & {12.63} & {7799} & {10.81} & {6705} & {12.07} & {4241}\\\hline
$4_{2}^{+}$ & {40.89} & {656} & {37.75} & {674} & {41.85} & {1514}\\\hline
$4_{3}^{+}$ & {47.26} & {127} & {44.17} & {83} & {49.03} & {20}\\\hline
$4_{4}^{+}$ & {x} & {x} & {48.01} & {1129} & {52.64} & {2011}\\\hline
$4_{5}^{+}$ & x & x & 53.68 & 67 & x & x\\\hline
\end{tabular}
\caption{Resonance energies and widths in the coupled channel calculation for
Minnesota (Mi), Volkov (V1) and modified Hasegawa-Nagata (MHN) potentials.
Resonance energies are in MeV, widths in keV.}%
\label{tab:Resonances}%
\end{table}

\section{Analysis}

The results of the previous section indicate an important effect of the
quadrupole and/or monopole deformation on the elastic $\alpha-\alpha$
continuum, appearing below the thresholds of the collective modes as
relatively narrow resonances above 30 MeV.

In order to analyze these results we examine the multi-channel wave function
for channel content, by calculating channel weights:%

\begin{equation}
W_{\upsilon,L}=\sum_{n}\left(  c_{nL}^{\upsilon}\right)  ^{2}%
\end{equation}

\begin{figure}
[ptb]
\begin{center}
\includegraphics[
natheight=9.541500in,
natwidth=7.047400in,
height=20.455cm,
width=15.1259cm
]%
{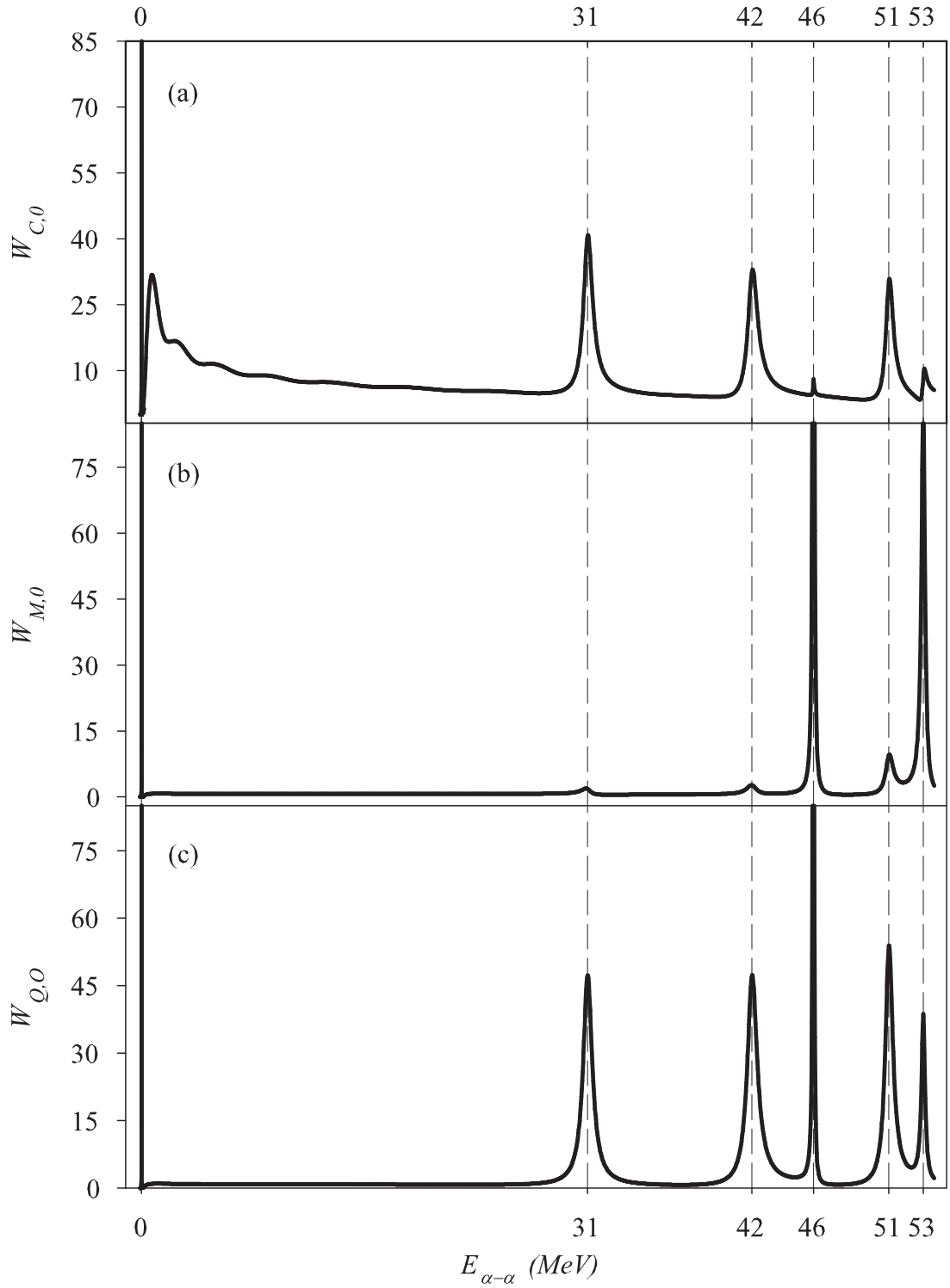}%
\caption{$L=0$ cluster (a), monopole (b) and quadrupole (c) channel weights
obtained with the Volkov force. $E_{\alpha-\alpha}$ is the c.m. energy with
respect to the cluster threshold.}%
\label{f:weights}%
\end{center}
\end{figure}

Figure \ref{f:weights} shows the channel weights for each of the three
channels ($\alpha-\alpha$, monopole and quadrupole) as a function of energy
for V1 and $L=0$. The figure includes the resonance positions for reference.
This picture indicates the strong polarization effects at the resonance
energies, emphasizing that the preferred configurations for $^{8}$Be in the
compound system at these resonance energies are dominated by the monopole
and/or quadrupole modes. Even more, the collective resonances suggest a
decoupled picture in the sense that each resonance is either essentially of a
quadrupole or monopole nature. It should be kept in mind however that the
channels are not orthogonal to each other, and that this fact blurs a
clear-cut comparison.

In order to further analyze the results in terms of the collective modes we
suggest to consider the orthogonal complement of the quadrupole and monopole
bases with respect to the $\alpha$-channel. This effectively removes the
non-orthogonality coupling between the $\alpha$-channel and the collective
ones. We denote the resulting subspaces by $\widetilde{Q}$, $\widetilde{M}$
and $\widetilde{QM}$, when the coupled orthogonal complement is used. The
transformed bases are easily obtained through Schmidt orthogonalization as%

\begin{gather}
\psi_{nL}^{\widetilde{M}}=\left[  \psi_{nL}^{M}-\left\langle \psi_{nL}%
^{M}|\psi_{nL}^{C}\right\rangle \psi_{nL}^{C}\right]  /\sqrt{1-\left\langle
\psi_{nL}^{M}|\psi_{nL}^{C}\right\rangle ^{2}};\\
\psi_{nL}^{\widetilde{Q}}=\left[  \psi_{nL}^{Q}-\left\langle \psi_{nL}%
^{Q}|\psi_{nL}^{C}\right\rangle \psi_{nL}^{C}\right]  /\sqrt{1-\left\langle
\psi_{nL}^{Q}|\psi_{nL}^{C}\right\rangle ^{2}}.\nonumber
\end{gather}
A further orthogonalization between the $\widetilde{Q}$ and $\widetilde{M}$
subspaces is not appropriate, as we will only be interested in either
quadrupole or monopole contributions for the characterization of resonance states.

\begin{figure}
[ptb]
\begin{center}
\includegraphics[
natheight=11.681000in,
natwidth=8.253800in,
height=21.3775cm,
width=15.1281cm
]%
{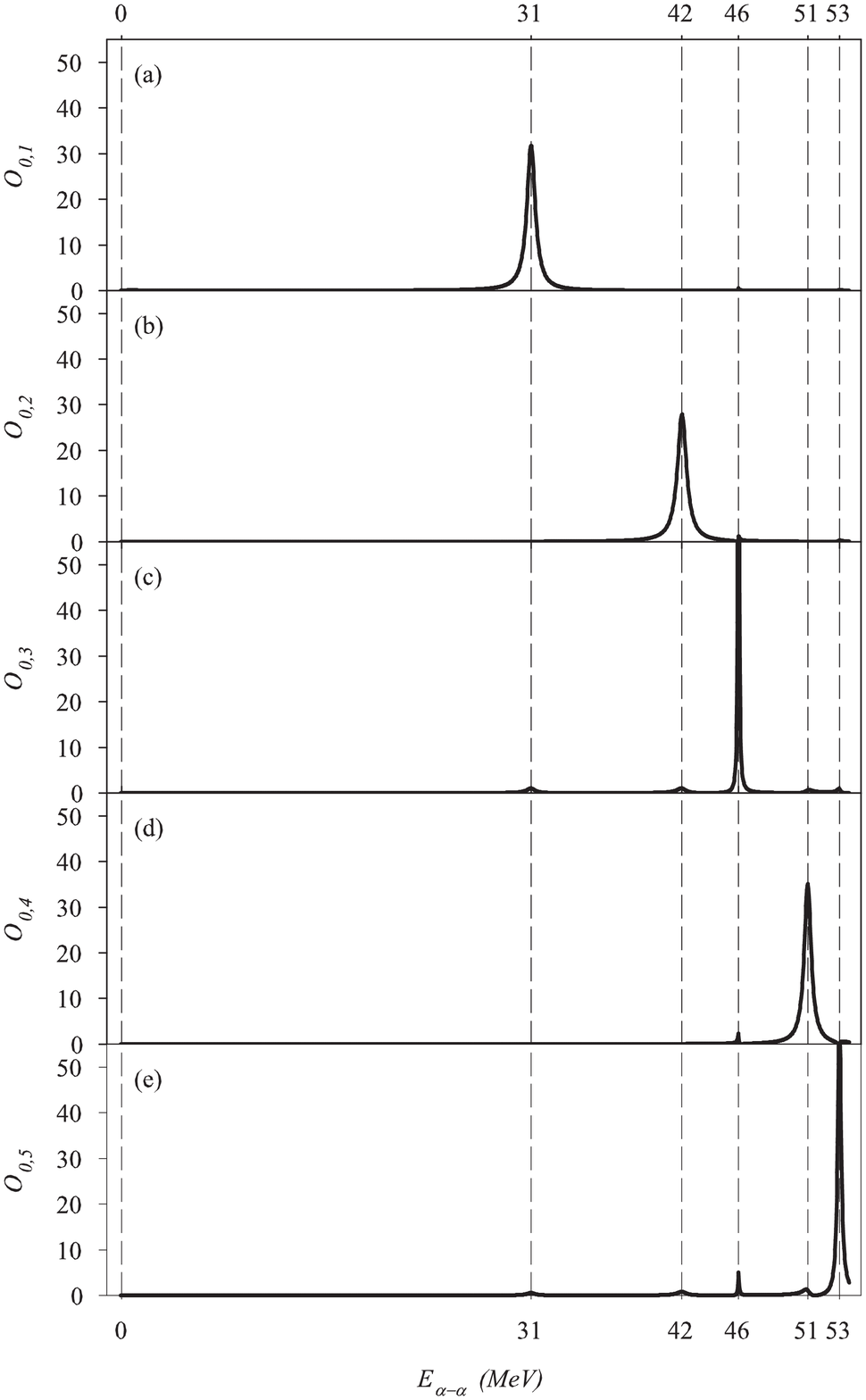}%
\caption{Overlap of $L=0$ energy wave function with the 1$^{\text{st}}$ (a), 2$^{\text{nd}}$ (b), 3$^{\text{rd}}$ (c), 4$^{\text{th}}$ (d) \ and 5$^{\text{th}}$ (e) eigenstates of the orthogonal complement to the cluster mode, calculated with the Volkov force. $E_{\alpha-\alpha}$ is the c.m.  energy with respect to the cluster threshold.}%
\label{f:orthch}%

\end{center}
\end{figure}

The multi-channel Hilbert subspace is now separated into two non-overlapping
parts, the open $\alpha$-channel and the closed orthogonal collective
components. This is similar to the Feshbach method \cite{kn:feshbach} of
projecting out the states of the \textquotedblleft external\textquotedblright%
\ decay channel (the cluster one) at continuum energy $E$, and constructing
the effective Hamiltonian in the \textquotedblleft internal\textquotedblright%
\ space of many body collective states. We calculate the bound eigenstates
$E_{iL}^{\widetilde{MQ}}$ in the orthogonal to the cluster channel subspace
and compare these to the resonance energies from the fully coupled scattering
calculation. The results for all potentials are summarized in Table
\ref{tab:content}. It is immediately apparent that the eigenenergies of the
coupled orthogonal subspace correspond almost exactly to the resonance
energies, except for the \textquotedblleft ground-state\textquotedblright\ one
which can be completely attributed to the Coulomb barrier in the
$\alpha-\alpha$ configuration. In Figure \ref{f:orthch} we show for V1 and
$L=0$ the content of the energy wave function in terms of the eigenstates of
the orthogonal complement ($O_{L,j}=\left\langle \psi_{L}^{E}|\psi
_{jL}^{\widetilde{MQ}}\right\rangle ^{2}$). It confirms the one-to-one
correspondence of the resonances to\ the orthogonal complement eigenstates.

A final aspect to be studied is the specific polarization nature of the
resonances suggested by Figure \ref{f:orthch}. To this end we calculate the
spectrum in both the $\widetilde{Q}$ and $\widetilde{M}$ spaces separately.
The combination of these uncoupled spectra is comparable to the coupled
$\widetilde{Q}$ and $\widetilde{M}$ diagonalization, indicating a limited
dynamical coupling between both subspaces. We indicate this in Table
\ref{tab:content}\ by showing for each eigenstate of the coupled orthogonal
complement the content ($\left\langle \psi_{iL}^{\widetilde{MQ}}|\psi
_{jL}^{\widetilde{M}}\right\rangle ^{2}$ or $\left\langle \psi_{iL}%
^{\widetilde{MQ}}|\psi_{jL}^{\widetilde{Q}}\right\rangle ^{2}$) of the most
prominent uncoupled eigenstate. This confirms in most cases an almost pure
polarization mode for each coupled eigenstate, and thus corresponding
resonance. These results are seen to be qualitatively identical for all three
potentials considered in this work, and for all $L=0,2$ and $4$ values.

\begin{table}[ptb]
\centering
\begin{tabular}
[c]{|l||c|c|c||c|c|c||c|c|c||}\hline
& \multicolumn{3}{c||}{Mi} & \multicolumn{3}{c||}{V1} &
\multicolumn{3}{c||}{MHN}\\\hline
& $E_{r}$ & $E_{iL}^{\widetilde{\text{MQ}}}$ & Mode & $E_{r}$ & $E_{i}%
^{\widetilde{\text{MQ}}}$ & Mode & $E_{r}$ & $E_{i}^{\widetilde{\text{MQ}}}$ &
Mode\\\hline
$0_{2}^{+}$ & {34.78} & {32.02} & {92\% $\widetilde{Q}_{1}$} & {30.51} &
{29.90} & {95\% $\widetilde{Q}_{1}$} & {34.13} & {33.78} & {97\%
$\widetilde{Q}_{1}$}\\\hline
$0_{3}^{+}$ & {44.45} & {44.19} & {75\% $\widetilde{Q}_{2}$} & {41.72} &
{41.52} & {90\% $\widetilde{Q}_{2}$} & {46.37} & {46.04} & {89\%
$\widetilde{Q}_{2}$}\\\hline
$0_{4}^{+}$ & {48.41} & {48.31} & {56\% $\widetilde{M}_{1}$} & {45.92} &
{45.77} & {43\% $\widetilde{M}_{1}$} & {51.71} & {51.42} & {66\%
$\widetilde{M}_{1}$}\\\hline
$0_{5}^{+}$ & x & x & x & {21.06} & {51.02} & {72\% $\widetilde{Q}_{3}$} &
{56.33} & {56.08} & {63\% $\widetilde{Q}_{3}$}\\\hline
$0_{6}^{+}$ & x & x & x & {53.41} & {53.33} & {63\% $\widetilde{M}_{2}$} & x &
x & x\\\hline
$2_{2}^{+}$ & {35.12} & {34.33} & {90\% $\widetilde{Q}_{1}$} & {32.59} &
{31.96} & {94\% $\widetilde{Q}_{1}$} & {36.41} & {36.08} & {96\%
$\widetilde{Q}_{1}$}\\\hline
$2_{3}^{+}$ & {46.51} & {46.21} & {66\% $\widetilde{Q}_{2}$} & {43.48} &
{43.27} & {93\% $\widetilde{Q}_{2}$} & {48.23} & {47.90} & {97\%
$\widetilde{Q}_{2}$}\\\hline
$2_{4}^{+}$ & {47.17} & {47.17} & {58\% $\widetilde{M}_{1}$} & {44.48} &
{44.46} & {66\% $\widetilde{M}_{1}$} & {50.63} & {49.69} & {71\%
$\widetilde{M}_{1}$}\\\hline
$2_{5}^{+}$ & x & x & x & {52.84} & {52.39} & {50\% $\widetilde{Q}_{3}$} &
{57.63} & {57.40} & {41\% $\widetilde{Q}_{3}$}\\\hline
$4_{2}^{+}$ & {40.89} & {39.89} & {80\% $\widetilde{Q}_{1}$} & {37.75} &
{36.94} & {88\% $\widetilde{Q}_{1}$} & {42.12} & {41.85} & {93\%
$\widetilde{Q}_{1}$}\\\hline
$4_{3}^{+}$ & {47.26} & {46.95} & {62\% $\widetilde{M}_{1}$} & {44.17} &
{43.84} & {71\% $\widetilde{M}_{1}$} & {49.52} & {49.03} & {78\%
$\widetilde{M}_{1}$}\\\hline
$4_{4}^{+}$ & x & x & x & {48.01} & {47.59} & {87\% $\widetilde{Q}_{2}$} &
{52.95} & {52.64} & {91\% $\widetilde{Q}_{2}$}\\\hline
$4_{5}^{+}$ & x & x & x & 53.68 & 53.60 & 86\% $\widetilde{M}_{2}$ & x & x &
x\\\hline
\end{tabular}
\caption{Comparison of the resonance energies $E^{\widetilde{MQ}}_{iL}$ in the
orthogonal complement to the cluster mode, denoted by $\widetilde{ \mbox{MQ}}%
$. All energies are in MeV. "Mode" stands for the content of the most
prominent orthogonal monopole $\psi_{jL}^{\widetilde{M}}$ or quadrupole
$\psi_{jL}^{\widetilde{M}}$ eigenstates in particular eigenstate $\psi
_{iL}^{\widetilde{MQ}}$.}%
\label{tab:content}%
\end{table}

From Table \ref{tab:content} one notices that the energy of the resonance
states lies above the corresponding eigenenergy of the collective orthogonal
complement. This confirms the results of ref. \cite{TwoChannProc} where it was
shown that the coupling between orthogonal open and closed channels transforms
bound state from the closed channel into a resonance with an energy above the
bound state one.

Table \ref{tab:content} also indicates a rotational behavior for the
quadrupole related resonances, closely following a $L(L+1)$ energy spacing,
whereas the monopole related resonances remain roughly at the same energy.

\section{Conclusions}

We have presented the results for the elastic $\alpha-\alpha$ spectrum for
$^{8}$Be obtained from a calculation in which the $\alpha$-cluster
configuration is coupled to the collective Sp(2,R) quadrupole and monopole
modes. The energy range considered is between the $\alpha$-channel and the
monopole (full 8 particle decay) thresholds. In order to accommodate both the
open $\alpha$-channel and the closed collective channels we considered a
multi-channel version of the Modified J-Matrix method.

The results indicate, apart from the well-known low-lying rotational band
attributed to Coulomb repulsion in the $\alpha-\alpha$ description, a rich
spectrum of relatively narrow resonances above 30 MeV. We have shown that the
resonances are connected to the eigenstates of the collective subspace,
orthogonalized to the open $\alpha$-channel. More specifically we have shown
that the resonances are essentially of quadrupole or monopole nature, and thus
exhibit a specific polarization of the nucleus. This indicates that both the
quadrupole and monopole eigenmodes remain mainly uncoupled as it was shown
earlier \cite{ar:caurierarickx-NPA398-467} for bound sd-nuclei. It shows that
both collective symmetries are important in the compound 8-particle system at
specific energies in $\alpha-\alpha$ scattering.

\section{Acknowledgments}

The authors A. Sytcheva and V. S. Vasilevsky are grateful to the Department of
Mathematics and Computer Science of the University of Antwerp (UA) for
hospitality. All authors acknowledge the financial support from the FWO-Vlaanderen.

%

%

%

%

\section{References}

\begin{thebibliography}{99}                                                                                               %


\bibitem {ar:ajzsel-NPA490-1}F.~Ajzenberg-Selove, Nucl. Phys. \textbf{A490},
1, (1988).

\bibitem {ar:tilley2004}D.~Tilley, J.~Kelley, J.~Godwin, D.~Millener,
J.~Purcell, C.~Sheu, and H.~Weller, Nucl. Phys. \textbf{A745}, 155 (2004).

\bibitem {ar:Bacher-PRL29-1331}A.~D. Bacher, F.~Resmini, H.~Conzett,
R.~de~Swiniarski, H.~Meiner, and J.~Ernst, Phys. Rev. Let. \textbf{29}, 1331 (1972).

\bibitem {AR:darriulat-PR-137-315}P.~Darriulat, G.~Igo, H.~Pugh, and
H.~Holmgren, Phys. Rev. \textbf{137}, 315 (1965).

\bibitem {AR:ostashko-JPG-20-1973}N.~Arena, I.~Barit, S.~Cavallaro,
A.~D'Arrigo, G.~Fazio, and G.~Giardina, J. Phys. G, \textbf{20}, 1973 (1994).

\bibitem {kn:wildtang77}K.~Wildermuth and Y.~Tang, \textit{A unified theory of
the nucleus}, (Braunschweig: Vieweg, 1977).

\bibitem {kn:tang81}Y.~Tang, \textit{Lecture Notes in Physics} \textbf{145},
(Berlin: Springer, 1981).

\bibitem {ar:Tang78}Y.~C. Tang, M.~LeMere, and D.~R. Thompson, Phys. Rep.
\textbf{47}, 167 (1978).

\bibitem {ar:wheeler-PR52-1083}J.~Wheeler, Phys. Rev. \textbf{52}, 1083 (1977).

\bibitem {ar:HillWheeler53}D.~Hill and J.~Wheeler, Phys. Rev. \textbf{89},
1102 (1953).

\bibitem {ar:saito77}S.~Saito, Prog. Theor. Phys. Suppl. \textbf{62}, 11 (1977).

\bibitem {ar:arai-PRC60-064315}K.~Arai and A.~T. Kruppa, Phys. Rev. C,
\textbf{60}, 064315 (1999).

\bibitem {AR:humblet-NPA-638-714}J.~Humblet, A.~Cs\'{o}t\'{o}, and
K.~Langanke, Nucl. Phys. \textbf{A638}, 714 (1998).

\bibitem {ar:caurierarickx-NPA398-467}F.~Arickx, J.~Broeckhove, E.~Caurier,
and P.~Van~Leuven, Nucl. Phys. \textbf{A398}, 467 (1982).

\bibitem {AR:deum-NPA-252-416}F.~Arickx, J.~Broeckhove, and E.~Deumens, Nucl.
Phys. \textbf{A269}, 318 (1979).

\bibitem {AR:arickx-NPA-284-264}F.~Arickx, Nucl. Phys. \textbf{A284}, 264 (1977).

\bibitem {ar:Rowe-RPP48-1419}D.~Rowe, Rep. Prog. Phys. \textbf{48}, 1419 (1985).

\bibitem {ar:suzuki-PTP75-1377}Y.~Suzuki, Prog. Theor. Phys. \textbf{75}, 1377 (1986).

\bibitem {ar:HechtBraunsch-NPA295-34}K.~Hecht and D.~Braunschweig, Nucl. Phys.
\textbf{A295}, 34 (1978).

\bibitem {ar:FilVasNes}G.~Filippov, V.~Vasilevsky, and A.~Nesterov, Nucl.
Phys. \textbf{A426}, 327 (1984).

\bibitem {ar:fil-NuovCim89}G.~Filippov, Nuov. Cim. \textbf{12}, 1 (1989).

\bibitem {kn:VV86collresE}G.~Filippov, V.~Vasilevsky, S.~Kruchinin, and
L.~Chopovsky, Sov. J. Nucl. Phys. \textbf{43}, 536 (1986).

\bibitem {ar:deumens:NPA423}E.~Deumens, Nucl. Phys. \textbf{A423}, 52 (1984).

\bibitem {AR:kruglanski-PRC-45-1321}D.~Baye and M.~Kruglanski, Phys. Rev. C
\textbf{45}, 1321 (1992).

\bibitem {ar:PRL88-10404}W.~Vanroose, J.~Broeckhove, and F.~Arickx, Phys. Rev.
Let. \textbf{88}, 10404 (2002).

\bibitem {ar:MJM4shortrange-jphysA-7769}J.~Broeckhove, F.~Arickx, W.~Vanroose,
and V.~Vasilevsky, J. Phys. A: Math. Gen, \textbf{37}, 7769 (2004).

\bibitem {ar:AM-AJP}F.~Arickx, J.~Broeckhove, P.~Van~Leuven, V.~Vasilevsky,
and G.~Filippov, Amer. J. Phys. \textbf{62}, 362 (1994).

\bibitem {kn:VA-PhysRev}V.~Vasilevsky and F.~Arickx, Phys. Rev. A~\textbf{55},
265 (1997).

\bibitem {ar:PRA-9-1974HY(JMM)}E.~J. Heller and H.~A. Yamani, Phys. Rev. A
\textbf{9}, 1201 (1974).

\bibitem {ar:volkov-NP74-33}A.~Volkov, Nucl. Phys. \textbf{74}, 33 (1965).

\bibitem {ar:tang-NPA286-53}D.~Thompson, M.~LeMere, and Y.~Tang, Nucl.
Phys.~\textbf{A286}, 53 (1977).

\bibitem {ar:tonabe-PTP53-677}F.~Tanabe, A.~Tohsaki, and R.~Tamagaki, Prog.
Theor. Phys.~\textbf{53}, 677 (1975).

\bibitem {ar:JPhysG18-1227}V.~Vasilevsky, G.~Filippov, F.~Arickx,
J.~Broeckhove, and P.~Van~Leuven, J. Phys. G: Nucl. Phys. \textbf{18}, 1227 (1992).

\bibitem {ar:vasnesarbr-PAN1997}V.~Vasilevsky, V.~Nesterov, F.~Arickx, and
J.~Broeckhove, Phys. Atom. Nuclei, \textbf{60}, 343 (1997).

\bibitem {ar:vas-PRC63-034606}V.~Vasilevsky, A.~Nesterov, F.~Arickx, and
J.~Broeckhove, Phys. Rev. C, \textbf{63}, 034606 (2001).

\bibitem {ar:vas-PRC63-034607}V.~Vasilevsky, A.~Nesterov, F.~Arickx, and
J.~Broeckhove, Phys. Rev. C, \textbf{63}, 034607 (2001).

\bibitem {ar:threecluster-alh}F.~Arickx, J.~Broeckhove, A.~Nesterov,
V.~Vasilvesky, and W.~Vanroose, \textit{J-matrix method and its applications},
A.~Alhaidari ed., (N.Y.: Nova Science Publishers, 2004, in press).

\bibitem {ar:hasegawa-PTP38-118}A.~Hasegawa and S.~Nagata, Prog. Theor. Phys.
\textbf{38}, 118 (1967).

\bibitem {ar:hasegawa-PTP45-1786}A.~Hasegawa and S.~Nagata, Prog. Theor. Phys.
\textbf{45}, 1786 (1971).

\bibitem {kn:feshbach}H.~Feshbach, \textit{Theoretical nuclear physics.
Nuclear reactions}, (USA: A Wiley-Interscience publication, 1991).

\bibitem {TwoChannProc}F.~Arickx, J.~Broeckhove, W.~Vanroose, P.~Van Leuven,
and V.~Vasilevsky, Proc. of the 6th. Int. Spring Seminar in Nuclear Physics
"Highlights of modern nuclear structure", St.Agata, May 1998, A.~Covello ed.,
(Singapore: World Scientific Publ. 1998), 485.
\end{thebibliography}
\end{document}